\documentclass[12pt]{article}
\pdfoutput=1
\usepackage{jhep-mod}
\usepackage{bm}
\usepackage{amssymb}
\usepackage{pifont}
\usepackage{slashed} 
\usepackage{slantsc} 
\usepackage[usenames,dvipsnames,svgnames,table]{xcolor}

\def\d{{\mathrm{d}}}

\title{%
The Hawking cascade from a black hole is extremely sparse
}
\author{Finnian Gray,}
\emailAdd{\\finnian.gray@msor.vuw.ac.nz}
\author{Sebastian Schuster,}
\emailAdd{sebastian.schuster@msor.vuw.ac.nz}
\author{Alexander Van--Brunt, \hbox{{\sf and}}}
\emailAdd{alexandervanbrunt@gmail.com}
\author{Matt Visser\,}
\emailAdd{matt.visser@msor.vuw.ac.nz}
\affiliation{School of Mathematics and Statistics,
Victoria University of Wellington; \\
PO Box 600, Wellington 6140, New Zealand.}
\abstract{\\
The Hawking flux from a black hole, (at least as seen from asymptotic infinity), is extremely sparse and thin, with the average time between emission of the successive Hawking quanta being many times larger than the natural timescale set by the energies of the emitted quanta.  
While this result has been known for over 30 years, it has largely been forgotten, possibly because many subsequent authors focussed mainly on the late-time high-temperature regime.  
We shall instead focus on the early-stage low-temperature regime, and shall both quantify and significantly extend these observations in a number of different ways. 
In particular we shall confront numerical estimates with semi-analytic approximations based on a naive Planck spectrum.

First we shall identify several natural dimensionless figures of merit, and thereby compare the mean time between emission of successive Hawking quanta to several distinct but quite natural timescales that can be associated with the emitted quanta, demonstrating that very large ratios are typical for emission of massless quanta from a Schwarzschild black hole. 
Furthermore these ratios are independent of the mass of the black hole as it slowly evolves. 
We shall then show that the situation for the more  general Reissner--Nordstr\"om and generic ``dirty'' black holes is even worse, at least as long as the surrounding matter  satisfies some suitable energy conditions. The situation for the  Kerr and Kerr--Newman black holes (or even for charged particle emission from a Reissner--Nordstr\"om black hole) is considerably trickier, and depends on a careful accounting of the super-radiant modes.

Overall, the Hawking quanta are seen to be dribbling out of the black hole one at a time, in an extremely slow cascade of 2-body decays.  Among other things, this implies that the Hawking flux is subject to ``shot noise''. 
Observationally, the Planck spectrum of the Hawking flux can only be determined by collecting and integrating data over a very long timescale.  
We conclude by connecting these points back to various kinematic aspects of the Hawking evaporation process. 

\medskip
\noindent
D{\sc{ate}}: 12 June 2015; 3 November 2015; \LaTeX-ed \today.
}
\enlargethispage{15pt}
\begin{document}
\maketitle
\clearpage
\section{Introduction}
\def\alert#1{{\color{red} #1 }}
\def\O{{\mathcal{O}}}

It is (or should be) well-known that the asymptotic Hawking flux from a black hole is extremely sparse and extremely thin. 
The interstitial gap, the average time between emission of successive Hawking quanta, is many times larger than the natural timescale set by the energies of the emitted quanta themselves.  
This result was established over 30 years ago~\cite{Page:1976a, Page:1976b, Page:1977, Page:thesis, Page:private}, but has largely been forgotten, quite possibly because many of the subsequent authors focussed mainly on the late-time high-temperature regime in the final stages of the evaporation process~\cite{Oliensis:1984, MacGibbon:1990, Halzen:1991, MacGibbon:2007, Page:2007, MacGibbon:2010}.  The early-stage low-temperature regime has recently been reconsidered by van Putten~\cite{van-Putten1, van-Putten2}.

We also shall focus on this early-stage low-temperature regime, and shall develop several simple semi-analytic estimates based on assuming an exact Planck spectrum, which is not the full story but is sufficient to give tolerable estimates, at least for spin-zero bosons. We shall compare these semi-analytic estimates with the numerical estimates of Page~\cite{Page:1976a,Page:1976b,Page:1977,Page:thesis}. (These  numerical estimates in particular include the effects of greybody factors.) Overall, and carefully separating out super-radiant contributions, (which in Hod's article~\cite{Hod} were simply lumped in with the Hawking effect), the sparsity of the Hawking flux is seen to persist throughout the entire evaporation process.

We shall carefully define and justify several natural dimensionless figures of merit, which are suitable for comparing the mean time between emission of successive Hawking quanta with several natural timescales that can be associated with the emitted quanta. 
We shall first focus on non-super-radiant situations:
We shall see that large ratios (very much greater than unity) are typical for emission of massless quanta from a Schwarzschild black hole. 
Furthermore these ratios are independent of the mass of the Schwarzschild black hole as it slowly evolves; certainly for as long as the Hawking temperature is well below the QCD scale. 
We shall then show that the situation for the more  general Reissner--Nordstr\"om and generic ``dirty'' black holes is even worse, at least as long as the surrounding matter  satisfies some suitable energy conditions, and as long as one is looking at the emission of uncharged Hawking quanta.
We shall also consider the effects of particle rest mass on the emitted Hawking quanta.

The onset of super-radiance considerably complicates the discussion. Super-radiance can (sometimes) occur for charged particle emission from Reissner--Nordstr\"om  black holes, and for a range of emitted quanta from rotating black holes. 
In particular the situation for the  Kerr and Kerr--Newman black holes is quite tricky, and depends on a careful accounting of the super-radiant modes. Certainly the quanta emitted in super-radiant modes are ``quantum vacuum radiation'', but whether or not one chooses to call them Hawking quanta is more problematic. 

\clearpage
Overall, throughout the entire history of the Hawking evaporation process, the (non-super-radiant) Hawking quanta will be seen to be dribbling out of the black hole one at a time, in an extremely slow cascade.  
Among other things, this implies the presence of ``shot noise'' in the Hawking flux. 
Observationally, the Planck-shaped spectrum of the Hawking flux can only be determined by collecting and integrating data over a very long time.
We shall conclude by connecting these points back to various kinematic aspects of the Hawking evaporation process, which is now seen to resemble a cascading chain of 2-body decay processes.

\section{Strategy}

We shall compare and contrast two approaches:
\begin{itemize}
\item 
As a zeroth-order approximation it should be perfectly adequate to  treat the Hawking flux as though it is simply Planck spectrum blackbody emission at the Hawking temperature. 
While we know that a  more careful treatment should at the very least include greybody, phase-space, and adiabaticity effects~\cite{Thermality}, 
(see also references~\cite{adiabatic1,adiabatic2} and~\cite{essential, observability}), nevertheless a zeroth order approximation using a blackbody spectrum should be quite sufficient to set the scale (if not the precise details) for the relevant issues we wish to consider. 

\item
At the  next order of approximation, the most significant limitation on naively treating the Hawking flux as Planckian blackbody emission emission comes from the greybody factors. Page~\cite{Page:1976a,Page:1976b,Page:1977,Page:thesis} resolves the Hawking flux into spin-dependent angular-momentum modes, and calculates various quantities of the form 
\begin{equation}
\langle Q \rangle =\sum_{\ell m} \int T_{s\ell m}(\omega) \; \langle n\rangle_\omega \; Q(\omega)\; {\d \omega}.
\end{equation}
Here $\langle n\rangle_\omega$ is a completely standard bosonic/fermionic occupation number, 
while the $T_{s\ell m}(\omega)$ are spin-dependent greybody factors, estimated by numerically solving the appropriate Regge--Wheeler/Zerilli equation for the radial waveform, this all being followed by a numerical integration over frequencies. 

Including the complications due to the integral over greybody factors does change the numerical value of our estimates, sometimes quite drastically, but does not change the qualitative nature of our results. (The much smaller effects due to adiabaticity and phase space constraints will remain negligible up to the final stages of the Hawking cascade when the black hole mass has shrunk down to  the Planck regime.) 
\end{itemize}
Between them, these two approaches give a good qualitative and quantitative handle on the sparsity of the Hawking flux. The blackbody emission approximation will be seen to work best for spin-zero, with higher spins seeing extra suppression (and increased sparsity) due to the angular momentum barrier.

\section{Flat space preliminaries}

The differential number flux, $(quanta)/(time)$, (of massless bosonic quanta emitted by a black body of temperature $T$, 
infinitesimal surface area $\d A$, and surface normal $\hat n$),  into a wave-number range $\d^3 \vec k$ is (in flat space) given by the utterly standard statistical mechanics result:
\begin{equation}
\d \Gamma =    {g\over(2\pi)^3}  \;  {c \; (\hat k \cdot \hat n) \over \exp(\hbar ck/k_BT)-1 }   \; \d^3 \vec k \; \d A.
\end{equation}
Here $g$ is the spin degeneracy factor; which is 1 for scalar bosons and 2 for massless bosons with non-zero spin.
Then integrating over azimuthal directions, 
\begin{equation}
\d \Gamma =    {g\over(2\pi)^3}  \;  {c \; 2\pi k^2 \cos\theta \over \exp(\hbar ck/k_BT)-1 }   \; \sin\theta \,\d\theta \, \d k \, \; \d A.
\end{equation}
Integrating over the remaining angle, $\theta\in(0,\pi/2)$, we see~\footnote{\, This angular integral is ultimately responsible for the ${1\over4}$ in the relationship between Stefan's constant and the radiation constant: $\sigma={1\over4}a c$. See \emph{e.g} Roberts~\cite{roberts}.}
\begin{equation}
\d \Gamma =    {g\over8\pi^2}  \;  {c k^2 \over \exp(\hbar ck/k_BT)-1 }   \; \d k \; \d A.
\end{equation}
The wave-number integral can easily be performed, 
so that for an object of finite surface area $A$ the total emitted number flux (see \emph{e.g.} Schwabl~\cite{Schwabl}) is:
\begin{equation}
\Gamma =      {g \,\zeta(3)\over4\pi^2}   \; {k_B^3 T^3\over\hbar^3 c^2} \; A.
\end{equation}
The reciprocal of this quantity, $\tau_\mathrm{gap} = 1/\Gamma$, 
is  the \emph{average} time interval between the emission of successive quanta. 

In counterpoint, the peak in the number spectrum occurs where $k^2/(e^{\hbar ck/k_BT}-1)$ is maximized, 
that is, at
\begin{equation}
\omega_\mathrm{\,peak\,number} = c k_\mathrm{\,peak\,number} =     {k_B T\over \hbar} \; \left( 2 + W(-2e^{-2})\right).
\end{equation}
Here $W(x)$ is the Lambert $W$-function, defined by $W(x) e^{W(x)} = x$. (See, for instance, references~\cite{Lambert1, Lambert2, primes}; 
the presence of the Lambert $W$ function in this calculation is not a ``deep'' result; 
it appears for exactly the same reason that the Lambert $W$-function appears in the constant characterizing Wien's displacement law.) 
The quanta emitted at this peak can only be temporally localized to within a few oscillation periods, 
so it is safe to take $\tau_\mathrm{localization} = 1/\nu_\mathrm{peak\,number}  = 2\pi/\omega_\mathrm{peak\,number}$ as a good estimate of the time required 
for each individual quantum to be emitted.~\footnote{\,Originally we had used $\tau_\mathrm{localization} = 1/\omega_\mathrm{peak\,number}$. In counterpoint,  Hod, see reference~\cite{Hod}, prefers to use $\tau_\mathrm{localization} = 1/\nu_\mathrm{peak\,number} = 2\pi/\omega_\mathrm{peak\,number}$. This numerical factor does not qualitatively change our results, but in the interests of being as conservative as possible we shall include the $2\pi$.}
Let us now define the dimensionless figure of merit
\begin{equation}
\eta_\mathrm{\,peak\,number}= {\tau_\mathrm{\,gap}\over\tau_\mathrm{\,localization}} =  {\nu_\mathrm{\,peak\,number}\over \Gamma } 
= {\pi\left( 2 + W(-2e^{-2})\right)\over g\; \zeta(3)} \; {\hbar^2 c^2\over k_B^2 T^2 A}.
\end{equation}
In terms of the so-called ``thermal wavelength'', $\lambda_\mathrm{\,thermal} = 2\pi \hbar c/(k_B T)$, this is
\begin{equation}
\eta_\mathrm{\,peak\,number} 
= {\left( 2 + W(-2e^{-2})\right)\over4 \pi g \;\zeta(3)} \; {\lambda^2_\mathrm{\,thermal}\over A}.
\end{equation}
If instead we consider the peak in the energy flux, rather than the peak in the number flux,
then the only change is that now the factor $\left( 2 + W(-2e^{-2})\right)\to\left( 3 + W(-3e^{-3})\right)$, and
\begin{equation}
\omega_\mathrm{\,peak\,energy} = c k_\mathrm{\,peak\,energy} =     {k_B T\over \hbar} \; \left( 3 + W(-3e^{-3})\right).
\end{equation}
Then we have
\begin{equation}
\eta_\mathrm{\,peak\,energy} 
= {\left( 3 + W(-3e^{-3})\right)\over4 \pi g \;\zeta(3)} \; {\lambda^2_\mathrm{\,thermal}\over A}.
\end{equation}
Similarly, we could use the average frequency to set the localisation timescale
\begin{equation}
\langle \omega \rangle  = {\int ck (\d \Gamma/\d k) \d k\over  \int (\d \Gamma/\d k)  \d k } = {\pi^4\over30\;\zeta(3)}\; {k_B T\over\hbar}. 
\end{equation}
The net result is that  $\left( 2 + W(-2e^{-2})\right)\to\pi^4/(30 \, \zeta(3))$, and so
\begin{equation}
\eta_\mathrm{\,average\,energy} 
= {\pi^2\over120 g \;\zeta(3)^2} \; {\lambda^2_\mathrm{\,thermal}\over A}.
\end{equation}
Now consider something more subtle;  let us divide the spectrum into ``wave-number bins'' and  take
\begin{equation}
\eta_\mathrm{\,binned} =  {1\over \displaystyle \int {2\pi\over c k} \; {\d\Gamma\over\d k}\;  \d k}.
\end{equation}
This quantity effectively calculates the decay rate into wave-number bins, of width $\d k$, centred on $\omega= c k$, 
compares this with the frequency $\nu = \omega/(2\pi)=ck/(2\pi)$, and then sums over all bins. 
A brief calculation yields
\begin{equation}
\eta_\mathrm{\,binned} = {24\over 2\pi g} \;{\hbar^2 c^2 \over k_B^2 T^2 A} =  {24\over 8\pi^3 g} {\lambda^2_\mathrm{\,thermal}\over A}.
\end{equation}
All of these sparsity estimates (in flat Minkowski space for now) take the form
\begin{equation}
\eta = \hbox{(dimensionless number)} \; {\lambda^2_\mathrm{\,thermal}\over g \, A}.
\end{equation}
Let us now introduce key aspects of black hole physics, adapting the discussion above to see how far we can get.

\section{Non-super-radiant situations}
\subsection{Schwarzschild black holes}

Under normal laboratory (and astronomical) conditions one is dealing with emitters whose surface area is extremely large in terms of the thermal wavelength, so in those situations  $\eta \ll 1$. 
However, this is exactly what fails for a Schwarzschild black hole.
\begin{itemize}
\item 
First $T\to T_H$, and for the Hawking temperature we have
\begin{equation}
k_B T_H =  {\hbar c\over 4\pi \; r_H}; \qquad \lambda_\mathrm{\,thermal} =  8 \pi^2 \; r_H.
\end{equation}
Note that the thermal wavelength is a factor $8\pi^2 \approx 78.95 \approx 80$ times larger than the Schwarzschild radius.
\item
Second, a subtlety arises here as to which ``area'' to use. Naively one might use $A\to A_H = 4\pi r_H^2$, but this corresponds to a cross section of ${1\over4} A_H = \pi r_H^2$, which is really only appropriate for some particle species in the low-frequency limit. At high frequencies (the ray optics limit) the cross section is universally given by $ {27\over4} \pi r_H^2 = {27\over16} A_H$~\cite{Page:1976a, MTW}. This is enhanced by a factor of ${27\over4}$,  and implies that,  (to smoothly match high frequency results), we should set $A\to A_\mathrm{\,effective} =  {27\over4} A_H = 27 \pi r_H^2$.~\footnote{\,This numerical factor does not qualitatively change our results, but in the interests of being as conservative as possible we shall include the ${27\over4}$.}
\end{itemize}

\subsubsection{Massless Bosons}

With these substitutions, for a Schwarzschild black hole we have
\begin{equation}
 {\lambda^2_\mathrm{\,thermal}\over A_\mathrm{\,effective}} = {64\pi^3\over 27} \approx 73.49635955... \gg 1,
\end{equation}
which is certainly much larger than unity.
Consequently
\begin{equation}
\eta_\mathrm{\,peak\, number} =  {32\pi^2\left( 2 + W(-2e^{-2})\right)\over27g\zeta(3)} = {15.50768123...\over g} \gg 1.
\end{equation}
As promised, the gap between successive Hawking quanta is on average much larger than the natural timescale associated with each individual emitted quantum. 
Note that this is a physical situation where the relevant dimensionless constant is not ``natural'' --- \emph{physically} it is not of order unity, though \emph{mathematically} one would quite legitimately still say the factor is $\O(1)$; neither is it zero nor infinity ---  the various numerical factors are important in determining the order of magnitude. 
Physically even more important is the fact that the mass of the Schwarzschild black hole drops out of the calculation, so that this calculation will be relevant as long as the dominant Hawking emission is into massless quanta.~\footnote{
We shall soon see that for other more general black holes the Schwarzschild result provides a mass-independent \emph{lower bound}.
}
This certainly holds throughout early stages of the evaporation where $k_B T_H \ll m_e c^2$, and will plausibly remain relevant until $k_B T_H \lesssim \Lambda_\mathrm{\,QCD}$. 

Similar calculations apply for the  other options we had considered for the localization timescale. Still working with the Schwarzschild black hole, we see that:
\begin{itemize}
\item 
If we consider the peak in the energy flux, rather than the peak in the number flux, then
\begin{equation}
\eta_\mathrm{\,peak\, energy} =  {32\pi^2\left( 3 + W(-3e^{-3})\right)\over27g\zeta(3)} = {27.45564528...\over g} \gg 1.
\end{equation}
\item
If we consider the average frequency then 
\begin{equation}
\eta_\mathrm{\,average\,energy} = {26.28537289...\over g}.
\end{equation}

\item
For the binned version of the $\eta$ parameter we have 
\begin{equation}
\eta_\mathrm{\,binned}= {14.22222222...\over g}.
\end{equation}
\end{itemize}
However one chooses the precise details to set up the calculation, and whatever the precise definition of $\eta$, 
it is clear that the time interval between successive emitted Hawking quanta, is on average, 
large compared to the natural timescale associated with the energy of the individual emitted quanta. 

We now compare this with numerical estimates along the lines of Page's results from the mid 1970's~\cite{Page:1976a,Page:1976b,Page:1977,Page:thesis}. The specific numbers will change, but the qualitative behaviour stays the same. Slightly modifying Page's 1976 analysis~\cite{Page:1976a,Page:1976b}, for emission of a spin $s$ quantum we define
\begin{equation}
\Gamma = \int T_{s\ell m}(\omega) \; \langle n\rangle_\omega \; {\d \omega};
\end{equation}
\begin{equation}
\langle \omega \rangle = \int \omega\; T_{s\ell m}(\omega) \; \langle n\rangle_\omega \; {\d \omega}/\Gamma;
\end{equation}
and
\begin{equation}
{1\over\eta_\mathrm{\,binned}} = 2\pi \sum_{\ell m} \int T_{s\ell m}(\omega) \; \langle n\rangle_\omega \; {\d \omega \over\omega}.
\end{equation}
Where available, we have used and adapted Page's 1976 numerical results. Where not otherwise available, we have  numerically estimated the greybody factors using product integral techniques adapted to the Regge--Wheeler/Zerilli potentials~\cite{to-appear}, followed by a numerical integration over frequencies. Our results are summarized in Table \ref{T:1}.

\begin{table}[htdp]
\caption{Semi-analytic and numerical estimates of $\eta$ for massless bosons  emitted from a Schwarzschild black hole. The semi-analytic estimates are from the current discussion. The numerical estimates  are from Page's 1976 results and our own extrapolations thereof. }
\begin{center}
\begin{tabular}{||c||c||c|c|c|c||}
\hline
\hline
BOSONIC &  g & $\eta_\mathrm{\,peak~number}$ & $\eta_\mathrm{\,peak~energy}$ & $\eta_\mathrm{\,average~frequency}$ &
$\eta_\mathrm{\,binned}$ \\
\hline
\hline
Semi-analytic & 1& ${32\pi^2\left( 2 + W(-2e^{-2})\right)\over27\zeta(3)}$ & ${32\pi^2\left( 3 + W(-3e^{-3})\right)\over27\zeta(3)}$ & 
${16\pi^6\over405\zeta(3)^2}$& ${128\over 9}$
\\
\hline
Value & 1& 15.508&27.465 & 26.285& 14.222 
\\
\hline\hline \hline
 $s=0$ & 1& 20.65& 27.83& 26.78& 16.31 
 \\
 \hline 
 $s=1$ & 2 &246.1& 259.1 & 244.5& 216.3 
 \\
 \hline 
 $s=2$ & 2 & 5076& 5219&4964 & 4692
 \\
\hline\hline
\end{tabular}
\end{center}
\label{T:1}
\end{table}%

\enlargethispage{40pt}

Overall we see that introducing the greybody factors, (because they suppress $\Gamma$), always have the effect of \emph{increasing} $\eta$. In particular, as one goes to higher spin there is a larger ``angular momentum barrier'' to overcome and the numerically estimated values of $\eta$ greatly exceed those obtained from the semi-analytic estimates based on a purely Planckian spectrum. For spin zero the semi-analytic estimates are in good agreement with the numerical results. For higher spin the semi-analytic estimates provide a lower bound, but generally sparsity of the Hawking flux is even more extreme than suggested by the semi-analytic estimate.

\subsubsection{Massless Fermions }

The modifications for massless fermions are straightforward. First we note that the differential emission rate changes to
 \begin{equation}
\d \Gamma =    {g\over16\pi^2}  \;  {c k^2 \over \exp(\hbar ck/k_B)+1 }   \; \d k \; \d A.
\end{equation}
(Note that the original version of the standard model of particle physics contained only massless chiral neutrinos with $g=1$; as soon as one extends the standard model to contain non-chiral neutrinos then $g=2$. We shall explicitly retain $g$, both for historical reasons and to facilitate comparisons with bosonic and Boltzmann results.)
Integrating over wavenumber and area
\begin{equation}
\Gamma =      {3 \,g\,\zeta(3)\over16\pi^2}   \; {k_B^3 T^3\over\hbar^3 c^2} \; A.
\end{equation}
For fixed $g$ this is certainly less than the bosonic result, by a factor $3/4$. Furthermore the peak in the number spectrum shifts upwards and is now at 
\begin{equation}
\omega_\mathrm{\,peak\,number} = c k_\mathrm{\,peak\,number} =     {k_B T\over \hbar} \; \left( 2 + W(2e^{-2})\right),
\end{equation}
while
\begin{equation}
\omega_\mathrm{\,peak\,energy} = c k_\mathrm{\,peak\,energy} =     {k_B T\over \hbar} \; \left( 3 + W(3e^{-3})\right).
\end{equation}
Including both effects, for a Schwarzschild black hole, (where,  as for the bosonic case, $A=27\pi r_H^2$ and $k_B T_H = \hbar c/(4\pi r_H)$), we have
\begin{equation}
\eta_\mathrm{\,peak\, number} =  {128\pi^2\left( 2 + W(+2e^{-2})\right)\over81 g\zeta(3)} = {28.77434355... \over g} \gg 1,
\end{equation}
and
\begin{equation}
\eta_\mathrm{\,peak\,energy} =  {128\pi^2\left( 3 + W(+3e^{-3})\right)\over81 g\zeta(3)} = {40.62426089...\over g} \gg 1.
\end{equation}
Using the average frequency
\begin{equation}
\langle \omega \rangle  \to {7\pi^4\over180\;\zeta(3)}\; {k_B T_H\over\hbar},
\end{equation}
and so
\begin{equation}
\eta_\mathrm{\,average~energy} = {224\pi^6\over3645\zeta(3)^2} = {40.88835782...\over g} \gg 1.
\end{equation}
Using the binned $\eta$ yields
\begin{equation}
\eta_\mathrm{\,binned} =   {256\over 9g} = {28.44444444...\over g} \gg 1.
\end{equation}
Thus it is clear that the time interval between emitted (massless fermionic) Hawking quanta is,  on average, 
large compared to the natural timescale associated with the energy of the emitted quanta. 
We again compare this, where we can, with numerical results including the effects of greybody factors. See Table \ref{T:2}.
Overall we see that the semi-analytic estimate is reasonably good for spin 1/2, though we again expect that for higher spin the angular momentum barrier will depress $\Gamma$ and enhance $\eta$.

\begin{table}[!htdp]
\caption{Semi-analytic and numerical estimates of $\eta$ for massless fermions  emitted from a Schwarzschild black hole. 
The semi-analytic estimates are from the current discussion. The numerical estimates  are from Page's 1976 results and our own extrapolations thereof. }
\begin{center}
\begin{tabular}{||c||c||c|c|c|c||}
\hline
\hline
FERMIONIC &  g & $\eta_\mathrm{\,peak~number}$ & $\eta_\mathrm{\,peak~energy}$ & $\eta_\mathrm{\,average~frequency}$ &
$\eta_\mathrm{\,binned}$ \\
\hline
\hline
Semi-analytic & g& ${128\pi^2\left( 2 + W(2e^{-2})\right)\over81g\zeta(3)}$ & ${128\pi^2\left( 3 + W(3e^{-3})\right)\over81g\zeta(3)}$ & 
${224\pi^6\over3645g\zeta(3)^2}$& ${256\over 9g}$
\\
\hline
Value & g& $28.773/g$&$40.624/g$ & $40.88/g$ & $28.444/g$
\\
\hline
Value & 2& 14.387&20.312 & 20.444& 14.222 
\\
\hline\hline \hline
 $s=1/2$ & 2&--- & 29.3& 27.6 & ---
 \\
\hline\hline
\end{tabular}
\end{center}
\label{T:2}
\end{table}%

\subsubsection{Massless Boltzmann particles }

For completeness, and for subsequent use, we now consider the case of particles which satisfy Boltzmann statistics. The differential emission rate changes to
 \begin{equation}
\d \Gamma =    {g\over8\pi^2}  \;  c k^2 \;  \exp(-\hbar ck/k_BT)  \; \d k \; \d A.
\end{equation}
A key observation is
\begin{equation}
 {1\over e^x-1} > {1\over e^x} > {1\over e^x + 1}
\end{equation}
This implies (for fixed $g$) that $\Gamma_\mathrm{\,bosonic} > \Gamma_\mathrm{\,boltzmann} > \Gamma_\mathrm{\,fermionic}$. 
In fact, integrating over wavenumber and area
\begin{equation}
\Gamma =      {g\over4\pi^2}   \; {k_B^3 T^3\over\hbar^3 c^2} \; A,
\end{equation}
which indeed lies between the bosonic and fermionic results. Furthermore the peak in the number spectrum also  lies between the bosonic and fermionic results and is now at 
\begin{equation}
\omega_\mathrm{\,peak\,number} = c k_\mathrm{\,peak\,number} =     { 2 k_B T_H\over \hbar},
\end{equation}
and
\begin{equation}
\omega_\mathrm{\,peak\,energy} = c k_\mathrm{\,peak\,energy} =     { 3 k_B T_H\over \hbar}.
\end{equation}
In this case the ordering is  $\omega_\mathrm{\,peak\,bosonic} <\omega_\mathrm{\,peak\,boltzmann} < \omega_\mathrm{\,peak\,fermionic}$ implying
\begin{equation}
\eta_\mathrm{\,bosonic} < \eta_\mathrm{\,boltzmann} < \eta_\mathrm{\,fermionic}.
\end{equation}
After a bit of work, for a Schwarzschild black hole we have
\begin{equation}
\eta_\mathrm{\,peak\,number} = {64\pi^2\over27 g} = {23.39461784...\over g}.
\end{equation}
\begin{equation}
\eta_\mathrm{\,peak\,energy} = {32\pi^2\over9 g} = {35.09192677...\over g}.
\end{equation}
For the Boltzmann spectrum, the average energy is at the peak energy, so
\begin{equation}
\eta_\mathrm{\,average\,energy} = {32\pi^2\over9 g} = {35.09192677...\over g}.
\end{equation}
Finally, the binned value of $\eta$ in this situation agrees with that obtained from the peak in the number spectrum
\begin{equation}
\eta_\mathrm{\,binned} = {64\pi^2\over27 g} = {23.39461784...\over g}.
\end{equation}
We summarize our results in Table \ref{T3}.  Overall we see that the semi-analytic estimates for the sparsity of the Boltzmann flux fits nicely between those for bosonic and fermionic fluxes.

\begin{table}[htdp]
\caption{Semi-analytic estimates of $\eta$ for massless Boltzmann particles emitted from a Schwarzschild black hole. These semi-analytic estimates are from the current discussion.}
\begin{center}
\begin{tabular}{|| c ||c|c|c|c|c||}
\hline
\hline
BOLTZMANN & g & $\eta_\mathrm{\,peak~number}$  & $\eta_\mathrm{\,peak~energy}$ & $\eta_\mathrm{\,average~frequency}$ &
$\eta_\mathrm{\,binned}$ \\
\hline
\hline
Semi-analytic &g &$  {64\pi^2\over27 g}  $ & ${32\pi^2\over9 g}$ & ${32\pi^2\over9 g}$ &  ${64\pi^2\over27 g}  $
\\
\hline
Value &g  & $23.395/g$ & $35.092/g$ &$35.092/g$ & $23.395/g$
\\
\hline\hline 

\end{tabular}
\end{center}
\label{T3}
\end{table}%

\subsubsection{The situation so far}

Up to this point we have only considered the emission of massless  Hawking quanta from Schwarzschild black holes. 
We have seen that the semi-analytic estimate (based on assuming an exactly Planckian spectrum for the emission) gives a reasonably accurate slight  under-estimate for the sparsity of $s=0$ and $s={1\over2}$ emission, but that for higher spin the angular momentum barrier quickly drives the sparsity parameter $\eta$ even higher. For photons numerical estimates including the effect of greybody factors yield $\eta > 200$, for gravitons $\eta> 4500$.

\subsection{Reissner--Nordstr\"om black holes (uncharged quanta)}

In all the situations we have considered so far we have $\eta\propto 1/(T_H^2 A_H)$; and we have then used properties of the Schwarzschild black hole to evaluate this quantity. For the more general Reissner--Nordstr\"om black holes, as long as we work in terms of the radius of the  inner and outer horizons, $r_\pm$, we can generically write
\begin{equation}
\kappa = {r_+-r_-\over 2r_+^2};  \qquad  k_B T_H = {\hbar c \kappa\over2\pi};\qquad A_H = 4\pi r_+^2.
\end{equation}
Consequently for the total emission of massless quanta, (which are then automatically guaranteed to be electrically neutral), and when working within the blackbody approximation,  we have:
\begin{equation}
\eta_{\hbox{\tiny Reissner--Nordstr\"om}} = \eta_{\hbox{\tiny Schwarzschild}} \times  {r_+^2\over(r_+-r_-)^2} \;\geq\; 
\eta_{\hbox{\tiny Schwarzschild}}.
\end{equation}
The key point here is that adding charge to the black hole serves only to make the (semi-analytic estimate for the) Hawking flux even more sparse.

\subsection{Dirty black holes}

So-called ``dirty black holes'' are black holes surrounded by some generic matter fields~\cite{dbh}. Let us keep everything static and spherically symmetric and write
\begin{equation}
\d s^2 = - e^{-2\phi(r)} (1-2m(r)/r) c^2 \d t^2 + {\d r^2\over 1-2m(r)/r} + r^2 (\d\theta^2 +\sin^2\theta \; \d\varphi^2).
\end{equation}
It is then straightforward to calculate the surface gravity and determine~\cite{dbh}
\begin{equation}
\kappa = {e^{-\phi_H}\over 2 r_H} \; \left(1-{8\pi G_N \rho_H r_H^2\over c^4}\right); \qquad A_H = 4\pi\, r_H^2. 
\end{equation}
(Here $\rho_H$ is the energy density at the horizon.)
Thence
\begin{equation}
\eta = \eta_{\mathrm{\,Schwarzschild}} \times { e^{\phi_H} \over 1-8\pi G_N\rho_H r_H^2/c^4}.
\end{equation}
But via the Einstein equations~\cite{dbh} 
\begin{equation}
\phi_H = {4\pi G_N\over c^4} \; \int_{r_H}^\infty {(\rho-p_r) \; r \over 1-2m(r)/r} \; \d r.
\end{equation}
Now as long as the null energy condition [NEC] holds in the radial direction, then  this quantity is guaranteed non-negative. Furthermore as long as the weak energy condition [WEC] is satisfied at the horizon,  then we have $\rho_H\geq 0$. 
Subject to these two classical energy conditions holding~\cite{twilight, Visser:1999} we have $\eta_\mathrm{\,dirty\,black\,hole} \geq \eta_{\mathrm{\,Schwarzschild}}$. 
The key point here is that adding extra matter to the black hole environment serves only to make the (semi-analytic estimate for the) Hawking flux even more sparse.

\subsection{Particle rest mass}

Adding particle rest masses for the emitted quanta one now has
\begin{equation}
\d \Gamma =    {g\over8\pi^2}  \;  {c k^2 \over \exp\left(\hbar \sqrt{\omega_0^2+c^2 k^2} /k_BT_H\right) \mp 1 }   \; \d k \; \d A,
\end{equation}
where $\omega_0 = m_0 c^2/\hbar$ is the Compton frequency. Unfortunately the relevant integral for the total number flux is no longer explicitly evaluable, at least not in any convenient form, and the best one can easily say is this:
\begin{equation}
\Gamma =      {g\over4\pi^2}   \; {k_B^3 T_H^3\over\hbar^3 c^2} \; A_H\times 
\left|\mathrm{Li}_3(\pm 1)\right| \times f(\hbar\omega_0/k_B T),
\end{equation}
where $\mathrm{Li}_n(z)$ denotes the polylog function, and~\footnote{
An explicit representation of the integral that appears in the numerator of $f(z)$ can be obtained in terms of an infinite sum of Bessel functions, but that expression does not seem to be particularly useful.
} 
\begin{equation}
f(z) = {\int_0^\infty  {x^2 \over \exp\left(\sqrt{z^2+x^2}\right) \mp 1 }  \; \d x\over \int_0^\infty  {x^2 \over \exp\left(x\right) \mp 1 } \; \d x}  \leq 1.
\end{equation}
Thus $\Gamma$ certainly decreases.

Similarly the location of the flux peak is no longer explicitly evaluable in closed form, (not even with the aid of the Lambert $W$-function), although it is possible to deduce on quite general grounds that the location of the flux peak moves upwards.  These two observations are enough, however, to guarantee that  $\eta$ increases as one introduces particle rest masses, and so the semi-analytic estimate for the Hawking flux again becomes sparser. 

In counterpoint, the binned version of $\eta$ becomes
\begin{equation}
{1\over\eta_\mathrm{\,binned}} =  2\pi \int {1\over \sqrt{\omega_0^2+c^2 k^2}} \; {\d\Gamma\over\d k}\;  \d k,
\end{equation}
whence, more explicitly~\footnote{
An explicit representation of this integral is also obtainable in terms of an infinite sum of Bessel functions, but this expression does not seem to be particularly useful.
} 
\begin{equation}
{1\over\eta_\mathrm{\,binned}} =  {g\over4\pi}  \;   \int   {1\over \sqrt{\omega_0^2+c^2 k^2}} \; 
 {c k^2\; A_H \over \exp\left(\hbar \sqrt{\omega_0^2+c^2 k^2} /k_BT_H\right) \mp 1 }   \; \d k.
\end{equation}
It is manifest from the above that $1/\eta_\mathrm{\,binned}$ again decreases, 
so we see that (the semi-analytic estimate for)  $\eta_\mathrm{\,binned}$ again increases, 
as particle rest masses are added to the emitted Hawking quanta. However, as we shall now see, once one enters the super-radiant regime the discussion becomes much messier.

\subsection{Summary}

At least for spherical symmetry, and in the non-super-radiant modes, the semi-analytic estimates for the sparsity of Hawking emission seem to be bounded below by that for the Schwarzschild black hole.  Insofar as we have been able to check the numerical estimates based on including the effects of greybody factors,  this continues to hold in more general situations; the Schwarzschild black hole provides a good lower bound on sparsity, at least in the absence of super-radiance.

\section{Super-radiant situations}

Super radiance can occur when for one reason or another one has a chemical potential $\mu$. The bosonic occupation number then becomes
\begin{equation}
\langle n\rangle_\omega \to {1\over \exp\{ (\hbar\omega-\mu)/k_B T_H\} - 1 }.
\end{equation}
If the chemical potential becomes positive ($\mu >0$) then the occupation number diverges for $\hbar\omega=\mu$ and becomes \emph{negative} for $\omega\in(0, \mu/\hbar)$.  In all situations of interest the greybody factor $T_{\ell sm}(\omega)$ simultaneously exhibits a zero and then becomes negative so that the product $T_{\ell sm}(\omega) \langle n\rangle_\omega$ remains finite and positive~\cite{Page:1976a}.  The region where both occupation number and greybody factor are negative is called the ``super-radiant regime''. This phenomenon occurs only for bosons,  and  (in the current black hole context) arises either due to charged particle emission from a charged black hole, or for low $L_z$ angular momentum modes from a rotating black hole (or a combination of these two effects).  For a more general discussion, not necessarily black-hole related, see for instance~\cite{Boonserm}, and references therein.

There is a definitional ambiguity as to whether one should include the super-radiant modes as part of the Hawking flux, or treat them separately.
For instance, super-radiance has a well-known classical limit --- the Penrose process for mining black hole rotational kinetic energy.  If one insists on considering the super-radiant modes as a subset of Hawking radiation, then one would also be forced to assert that the Penrose process is the classical limit of the Hawking process --- an assertion that many would feel is excessive in its generality. 

\enlargethispage{20pt}

Certainly both the super-radiant modes and non-super-radiant modes are each subsets of ``quantum vacuum radiation", 
but the underlying physics is very different and we feel there are good reasons to keep the two notions distinct. 
Hawking quanta are related to horizons, while super-radiance is related to (generalized) ergo-regions.  
Hawking quanta are temperature dominated, while super-radiant quanta are chemical potential dominated.
We shall reserve the phrase ``Hawking modes'' for the non-super-radiant subset of the quantum vacuum radiation modes.

\subsection{Reissner--Nordstr\"om black holes (charged quanta)}

The key change for the emission of charged quanta from a charged black hole is that the boseonic occupation factor picks up a ``chemical potential'' $\mu = q V_H$
 depending on the charge $q$ of the emitted quanta and the electromagnetic potential $V_H$ at the horizon:
\begin{equation}
\langle n\rangle_\omega \to {1\over \exp\{ (\hbar\omega-qV_H)/k_B T_H\} - 1 }
\end{equation}
If the mass of the emitted particle satisfies $m_0 c^2 < q V_H$ then there is a critical frequency at which the bosonic occupation number diverges 
(and below which the bosonic occupation number is actually negative).  
In that super-radiant region there is no escaping the relevance of the grey-body factors ---  the $T_{\ell mn}(\omega)$ non-perturbatively differ from unity, and there is no longer any good reason to really trust the semi-analytic approximation based on a pure Planck spectrum.  In this context, we can in fact extend the usual notion of a ergo-surface in a particle-dependent manner by 
defining the radius $r_E$ of the ergo-surface to be
\begin{equation}
m_0 c^2 = q V(r_E),
\end{equation}
while the associated ergo-region is defined by
\begin{equation}
\{ r : m_0 c^2 < q V(r) \}.
\end{equation}
Note that these generalized ergo-regions depend on both rest mass and electric charge.
Astrophysically these generalized ergo-regions are of limited usefulness, (since astrophysical black holes tend to quickly neutralise). 
Accordingly, we find it most useful to concentrate the discussion on the Kerr spacetime.

\subsection{Kerr and Kerr--Newman black holes}

For the Schwarzschild black hole in the pure Planck blackbody approximation we saw that $\eta\propto 1/(T_H^2 A_H)$; and we then used properties of the Schwarzschild black hole to evaluate this quantity. For the more general  Kerr and Kerr--Newman black holes, as long as we work in terms of the radius of the  inner and outer horizons, $r_\pm$, we can generically write
\begin{equation}
\kappa = {r_+-r_-\over 2(r_+^2+a^2)};  \qquad  k_B T_H = {\hbar c \kappa\over2\pi};\qquad A_H = 4\pi(r_+^2+a^2).
\end{equation}
Therefore, within the context of the semi-analytic approximation
\begin{equation}
(\kappa^2 A_H)_{\mathrm{\,Kerr}}  =  (\kappa^2 A_H)_\mathrm{\,Schwarzschild} \times {(r_+-r_-)^2\over r_+^2+a^2}.
\end{equation}
Consequently for the total emission of massless quanta, (which are then automatically guaranteed to be electrically neutral), and when working within the blackbody approximation,  we have:
\begin{equation}
\eta_\mathrm{\,Kerr} = \eta_{\mathrm{\,Schwarzschild}} \times  {r_+^2+a^2\over(r_+-r_-)^2} \;\geq\; \eta_{\mathrm{\,Schwarzschild}}.
\end{equation}
The key point here is that adding charge and/or angular momentum to the black hole serves only to make the  (semi-analytic estimate for the)  Hawking flux even more sparse.

In contrast, Hod~\cite{Hod} asserts the equivalent of $\eta = \O(1)$ for highly extremal Kerr. 
(That is, for $\hat a = J/M^2 = a/M \lesssim 1$.)
Hod achieves this by lumping super-radiance into the Hawking flux. 
In the extremal limit ($\kappa\to0$) the greybody factors are approximately~\cite{Page:1976a}
\begin{equation}
T(\omega) \approx C_{\ell, s} \; (A_H\; \omega \;[\omega-m\Omega_H])^{2\ell+1}.
\end{equation}
The bosonic occupation number is
\begin{equation}
\langle n\rangle_\omega = {1\over \exp\{\hbar(\omega-m\Omega_H)/k_B T_H\} -1 }.
\end{equation}
These both change sign at $\omega=m\Omega_H$. 
\begin{itemize}
\item
For $0 \leq \omega \leq m \Omega_H$ the greybody factor is negative.
\item
For $0 \leq \omega \leq m \Omega_H$ the bosonic occupation number is negative.
\end{itemize}
Consequently, for  $0 \leq \omega \leq m \Omega_H$ the super-radiant emission is not well-approximated by a blackbody.
So it makes sense to split the integral into two regions:
\begin{itemize}
\item $0 \leq \omega \leq m \Omega_H$ --- super-radiant emission.
\item $m \Omega_H \leq\omega\leq\infty$ --- Hawking emission.
\end{itemize}
It is well-known that  in the extremal limit ($\kappa\to0$) super-radiance is known to dominate over the Hawking flux~\cite{Page:1976a,Page:1976b,Page:1977,Page:thesis,Page:private}. 
If we concentrate on the binned version of $\eta$ (which for present purposes is more important than the flux) then slightly modifying Page's analysis
\begin{equation}
{1\over\eta} =2\pi \sum_{\ell m} \int T(\omega) \; \langle n\rangle_\omega \; {\d \omega \over\omega}.
\end{equation}
Then in the near-extremal limit:
\begin{equation}
{1\over\eta} 
\approx (A_H \; \Omega_H^2)^{2\ell+1} \sum_{\ell m}  m^2 C_{\ell, s} \int \;  {(x[x-1])^{2\ell+1}  \over \exp(\epsilon[x-1])-  1  }\; {\d x\over x} .
\end{equation}
Here:
 $x = \omega/(m \Omega_H)$ and 
$\epsilon = (\hbar m \Omega_H)/(k_B T_H) \gg\!\!\gg 1$. 
As usual, the emission is dominated by the lowest available angular momentum state $\ell=m=s$:
\begin{equation}
{1\over\eta} \approx C_{s, s} \; (A_H \; s^2 \; \Omega_H^2)^{2s+1}  \int \;  {(x[x-1])^{2s+1}  \over \exp(\epsilon[x-1])-  1  }\;  {\d x\over x} .
\end{equation}
Splitting into super-radiant and Hawking modes:~\footnote{%
Note the split takes place ``in parallel'';  that is
\[
{1\over\eta} = {1\over\eta_\mathrm{\,super\hbox{-}radiant}}  + {1\over\eta_\mathrm{\,Hawking}} .
\]
}
\begin{equation}
{1\over\eta_\mathrm{\,super\hbox{-}radiant}} \approx C_{s, s} \; (A_H s^2 \Omega_H^2)^{2s+1} \int_0^1 \;  {(x[x-1])^{2s+1}  \over \exp(\epsilon[x-1])-  1  }\;  {\d x\over x} ;
\end{equation}
and
\begin{equation}
{1\over\eta_\mathrm{\,Hawking}} \approx C_{s, s}\; (A_H s^2 \Omega_H^2)^{2s+1}  \int_1^\infty \;  {(x[x-1])^{2s+1}  \over \exp(\epsilon[x-1])-  1  }\;  {\d x\over x} .
\end{equation}
Recalling that $\epsilon  \gg\!\!\gg 1$,
some brute-force integration then yields the estimates:
\begin{equation}
{\eta_\mathrm{\,super\hbox{-}radiant}}  = \O\left(\epsilon^0\right); \qquad 
{\eta_\mathrm{\,Hawking}} = \O\left(\epsilon^{2s+2}\right) \gg\!\!\gg 1.
\end{equation}
So super-radiance dominates in the extremal limit.
In fact, super-radiance leads to rapid spin-down with small energy loss~\cite{Page:1976a, Page:1976b, Page:1977, Page:thesis, Page:private}, 
until the system goes non-super-radiant, and then the ``normal'' Hawking effect takes over.
The quantitative details are messy, but the overall message is clear: Sparsity of the Hawking flux is the dominant feature of the Hawking evaporation process.

\section{Discussion}

So long as the temperature remains below the rest energy of the electron we can safely ignore emission into charged particles. 
Indeed, considering emission into individual angular momentum modes makes sense only within the specific framework of the Page analysis~\cite{Page:1976a, Page:1976b, Page:1977}, and such complications can be safely ignored as long as the Hawking radiation can be treated as an approximate blackbody. 
Collecting all the various results we see that throughout the initial low-temperature stages of Hawking evaporation:
\begin{itemize}
\item For bosonic Hawking quanta $\eta \gg 1$ in all situations of interest.
\item For fermionic Hawking quanta $\eta \gg 1$ in all situations of interest.
\item For Boltzmann Hawking quanta $\eta \gg 1$ in all situations of interest.
\end{itemize}
This is best summarized as the statement that the Hawking flux is extremely sparse --- the average time between emission of Hawking quanta is very large compared to the timescale set by the energies of the Hawking quanta.  The Hawking quanta are dribbling out one-by-one, with very large interstitial gaps.  This phenomenon persists throughout the entire evolution of the black hole, both in early stages and (with suitable minor modifications) in late stages. (Earlier authors have largely confined their attention to the late-time high-temperature regime~\cite{Oliensis:1984, MacGibbon:1990, Halzen:1991, MacGibbon:2007, Page:2007, MacGibbon:2010}.) 

The bad news is this: Compared to numerics the semi-analytic estimates are often off (under-estimating sparsity) by factors of 100 or even more.
Sparsity if anything increases when you do numerics that includes the effects of greybody factors.
Why the discrepancy? 
Individual photons can at best be localized (both in the direction of motion and {transversely}) to about a wavelength.
The usual blackbody emission estimate is based on treating the photons as particles, not waves.
Once the physical size of the emitter becomes smaller than a ``typical wavelength", (more exactly, the thermal wavelength),
the usual blackbody emission estimate is no longer trustworthy, providing at best a qualitative estimate, typically a lower bound on sparsity.
But note that the usual statistical mechanics result
\begin{equation}
\d \Gamma =    {g\over8\pi^2}  \;  {c k^2 \over \exp(\hbar ck/k_BT_H)-1 }   \; \d k \; \d A.
\end{equation}
should still be valid for $\omega\to\infty$.
So semi-analytic estimates are OK in the Boltzmann tail,
but may mis-estimate things in the low-frequency regime.
Sparsity, however, is here to stay --- modulo arguments on how to classify super-radiance.

\bigskip
\noindent
The sparseness of the Hawking flux has a number of  perhaps unexpected kinematical implications:
\begin{itemize}
\item While early-stage Hawking radiation from Schwarzschild or Reissner--Nordstr\"om black holes is spherically symmetric, this spherical symmetry is only a long-term statistical statement obtained after averaging over very many Hawking quanta. 
\item
Early-stage Hawking evaporation should be seen as a long chain of independent 2-body decay processes involving photons, gravitons, and neutrinos. (Similarly, late-stage Hawking evaporation, once the temperature exceeds $\Lambda_\mathrm{\,QCD}$, should be viewed as a long chain of 2-body decay processes proceeding via the emission of hadronic jets~\cite{Oliensis:1984, MacGibbon:1990, Halzen:1991, MacGibbon:2007, Page:2007, MacGibbon:2010}.)
\item
When analyzing the emission of individual Hawking quanta one should use the special relativistic kinematics that is applicable in the asymptotic spatial region. Consider a black hole initially of rest energy $m_i c^2$ which emits (in its rest frame, but as seen from spatial infinity) a particle of energy $\hbar\omega$ and rest energy $\hbar\omega_0$, thereby reducing its rest energy to $m_f c^2$. Then the conservation of 4-momentum, (when written in terms of the 4-velocities $V$), implies
\begin{equation}
(m_i c^2) \; V_i = (m_f c^2) \; V_f + (\hbar \omega_0) \; V_H.
\end{equation}
Therefore
\begin{equation}
(m_i c^2)^2 + (\hbar\omega_0)^2 - 2 (m_ic^2) (\hbar\omega) = (m_f c^2)^2,
\end{equation}
and so
\begin{equation}
\hbar\omega =  {(m_i c^2)^2 +(\hbar\omega_0)^2 -  (m_f c^2)^2\over 2 m_i c^2}.
\end{equation}
Depending on whether or not one views black hole masses as being quantized or continuous, one can view this either as a normal 2-body decay, or as the decay of one IMP (``indefinite mass particle'' $\approx$ unparticle) into another IMP~\cite{IMP, IMP2}. It may be profitable to reconsider and reanalyze the entire Hawking evaporation process from this point of view.~\footnote{
MV's early views on the importance of the Hawking cascade can be found in reference~\cite{Visser:1992}.
 Although MV is no longer in favour of the particular way that black hole entropy was discretized  in that article, the comments regarding the importance of the final ``particle cascade'' leading to complete evaporation of Planck-scale black holes still hold.}%

\item
More radically, the fact that the Hawking flux is so sparse calls into question the appropriateness of the use of bosonic and fermionic statistics, (at least for non-equilibrium Hawking evaporation into the Unruh vacuum state). For the intrinsically equilibrium Hartle--Hawking vacuum there is absolutely no doubt that bosonic and fermionic statistics are the relevant ones, but when quanta are so well separated as they are in the non-equilibrium Hawking evaporation process it is less clear that bosonic and fermionic statistics make sense, since these normally require multiple occupation of the same quantum mode, typically achieved by being at (or at least temporarily near) equilibrium.~\footnote{
For instance, this certainly happens in a ``quasi-continuous-wave''  laser, where photon emission is not really continuous --- finite bandwidth is inversely related to finite coherence time. Typically one has short microsecond photon pulses with very high occupation numbers, so bosonic statistics is completely appropriate.
}
On the other hand, in the super-radiant regime it is clear that bosonic statistics is \emph{essential}, and that the blackbody approximation (Planck approximation) for the emitted flux is a poor one. 

\item
Finally, and even more speculatively, the sparsity of the Hawking flux provides a very obvious place to hide information without significant energy cost --- the gaps between the Hawking quanta can easily encode significant information without disturbing the (time-averaged) emission spectrum --- this would effectively be a ``phase modulated'' Hawking flux.
\end{itemize}
In summary: The extreme sparsity of the Hawking flux is a rather under-appreciated feature of the Hawking evaporation process. Hawking evaporation is seen to be a slow dribbling out of Hawking quanta, with the individual quanta being well separated in time. (The gap between individual quanta being hundreds of times longer than the natural timescales associated with the quanta.) This ``decay chain'' viewpoint has a number of tantalisingly interesting physical implications.

\acknowledgments

We wish to thank Don Page for his helpful comments and interest in this work.

\bigskip
\noindent
This research was supported by the Marsden Fund, through a grant administered by the Royal Society of New Zealand. 
\\
FG was also supported by a Victoria University of Wellington MSc Scholarship.
\\
SS was also supported by a Victoria University of Wellington PhD Scholarship.
\\
AVB was also supported by a Victoria University of Wellington Summer Scholarship.


\end{document}